\documentclass[aps,pra,twocolumn,superscriptaddress,showpacs,showkeys,amsmath,amssymb]{revtex4}

\usepackage{amsfonts}
\usepackage{mathrsfs}
\usepackage{latexsym}
\usepackage[cp1251]{inputenc}
\usepackage{graphicx}
\usepackage{dcolumn}
\usepackage{bm}
\usepackage{color}
\RequirePackage{ifthen}
\RequirePackage[pdfstartview=FitH]{hyperref}

\begin{document}

    \title{Condensation and superfluidity of $SU(N)$ Bose gas
    }
    \author{Orest~Hryhorchak}
    \author{Volodymyr~Pastukhov\footnote{e-mail: volodyapastukhov@gmail.com}}
    \affiliation{Department for Theoretical Physics, Ivan Franko National University of Lviv,\\ 12 Drahomanov Street, Lviv-5, 79005, Ukraine}

    \date{\today}



    \begin{abstract}
    We perform the comprehensive comparison of properties of the condensate and superfluid densities for the $N$-component three-dimensional Bose gas with the symmetric inter- and intraspecies short-range interaction between particles. In particular, based on the large-$N$ expansion approach for many-boson systems we obtain general expression for density of the superfluid component that at very low temperatures reproduce the well-know Landau's formula and non-trivially includes the thermal fluctuations in the finite-temperature region, and compare it to the condensate density calculated previously. The numerically evaluated temperature dependencies are in a qualitatively good agreement with the results of Monte Carlo simulations.

    \end{abstract}

    \maketitle

\section{Introduction}
\label{sec1} \setcounter{equation}{0}
Superfluidity is the first observable phenomenon that comes to mind when one talks about macroscopic low-temperature behavior of interacting many-boson systems. Being induced by quantum effects, discovered more than 80 years ago \cite{Kapitza,Allen} in liquid ${^4}$He and described \cite{Tisza,Landau} by means of the two-fluid concept, the superfluid behavior is a consequence of emergence of the off-diagonal (quasi)-long-range order in the Bose system \cite{Pitaevskii}. Particularly, for spatial dimensionalities larger than two it always goes together with the truly microscopic characteristic of bosonic systems, namely, the Bose-Einstein condensation (BEC) phenomenon. Therefore, it is generally believed that the occurrence of superfluidity should be associated with BEC transition in these systems. This notion is supported by experimental measurements \cite{Onofrio,Raman}, results of the Monte Carlo (MC) simulations \cite{Kashurnikov,Arnold,Prokofev,Nho} and analytical approaches \cite{Floerchinger,Isaule}, nonetheless, an interesting alternative points of view often appear \cite{Cooper_etal,Kruglov}.

Experimental realization of almost uniform BECs \cite{Gaunt} and recent measurements of the ground-state Bogoliubov condensate depletion \cite{Lopes_etal} in the homogeneous three-dimensional Bose gas together with the observation \cite{Ville} of sound propagation in the two-dimensional superfluids at finite temperatures allow for further experimental investigations of the interplay between quantum and thermal effects in the many-boson systems with continuous translational symmetry. From the theoretical point of view, the account of quantum and thermal fluctuations is fundamentally important for the description of these systems in the whole temperature range including both very low temperatures near absolute zero and a narrow region of critical point. The standard Bogoliubov prescription adopted for thermodynamics of a dilute Bose gas is capable to capture only the low-temperature properties of a system and in order to obtain the correct (at least on the qualitative level) behavior near the BEC transition temperature one has to use more sophisticated treatments. An extension of Bogoliubov's theory that naturally combines quantum and thermal fluctuations in the simplest fashion is given by the large-$N$ expansion approach \cite{Andersen,Chien,Diehl}, which was recently shown to predict a qualitatively correct shift of the critical temperature \cite{Hryhorchak} and thermodynamics \cite{Hryhorchak_19} for the Bose gas with a point-like two-body repulsive potential in a whole range of the interaction parameter. The objective of present study is to explore, by means of the $1/N$-expansion techniques, the superfluid properties of this system.

\section{Formulation}
The discussed model is described by the imaginary-time action \cite{Popov}, which after application of the Hubbard-Stratonovich transformation takes the following form 
\begin{eqnarray}\label{S}
	S=\int dx\, \psi^*_{\sigma}(x)\left\{\partial_{\tau}-\xi-i\varphi(x)\right\}
	\psi_{\sigma}(x)\nonumber\\
	-\frac{N}{2g}\int dx
	\varphi^2(x),
\end{eqnarray}
where the point-like two-body interaction potentials (both intra- and interspecies) between Bose particles is characterized by a single coupling constant $g/N$. The summation over repeating index $\sigma=1,\ldots,N$ is assumed.
Other notations are typical: $x\equiv(\tau, {\bf r})$, integration $\int
dx=\int_0^{1/T}d\tau\int_Vd{\bf r}$ is carried out over the $3+1$ `volume' $V/T$ (here $T$ is the temperature of the system), the $1/T$-periodic in imaginary-time variable $\tau$ complex fields $\psi_{\sigma}(x)$ describe $N$ species of bosons and the second-order differential operator $\xi=-\hbar^2\nabla^2/2m-\mu$ contains the chemical potential $\mu$ that fixes the total number of particles in a system.

In the low-temperature phase, the Bose condensate occurs. It means that the spatially uniform part $\psi_0$ of fields $\psi_{\sigma}(x)=\psi_0+\tilde{\psi}_{\sigma}(x)$
describing particle degree of freedom should be singled out. We also have to explicitly separate the saddle-point value $\varphi_0$ of the real field $\varphi(x)=\varphi_0+\tilde{\varphi}(x)$ that incorporate the collective behavior of the system. After that, the shifted complex fields $\tilde{\psi}_{\sigma}(x)$ can be integrated out to yield the following effective action
\begin{align}\label{S_eff}
& S_{{\rm eff}}=-\frac{NV\varphi_0^2}{2Tg}+NVn_0\tilde{\mu}/T\nonumber\\ 
&-Nn_0\int dx\, \tilde{\varphi}(x)\left\{\tilde{\xi}+i\tilde{\varphi}(x)-\partial_{\tau}\right\}^{-1}\tilde{\varphi}(x)\nonumber\\
&-\frac{N}{2g}\int dx
\tilde{\varphi}^2(x)-N{\rm Sp}\ln\left\{\tilde{\xi}+i\tilde{\varphi}(x)-\partial_{\tau}\right\},
\end{align}
where $\tilde{\xi}=\xi|_{\mu\to\tilde{\mu}}$, ($\tilde{\mu}=\mu-i\varphi_0$), $n_0=|\psi_0|^2$ is the Bose condensate density of each constituent and ${\rm Sp}\ldots$ denotes the trace of the appropriate differential operator. Equations $-\left(\partial \Omega/\partial\mu\right)_{n_0, \varphi_0}=NVn$, $\left(\partial \Omega/\partial
\varphi_0\right)_{\mu,n_0}=0$, $\left(\partial \Omega/\partial
n_0\right)_{\mu,\varphi_0}=0$ relate the Bose condensate density $n_0$, the chemical potential of the system and total particle number $NVn$ to non-zero expectation value of field $\varphi(x)$.

The further analysis trivializes in the large-$N$ limit. The main contribution, which is of order $N$, is now coming from the ideal-gas term, while the quadratic in fields $\tilde{\varphi}(x)$ part of action $S_{\rm eff}$, which impacts the terms of order unity in the thermodynamic potential, can be also exactly taken into account
\begin{eqnarray}\label{Omega}
\frac{\Omega}{N}=\frac{V\varphi_0^2}{2g}-Vn_0\tilde{\mu}-T\sum_{{\bf k}}\ln\left(1-e^{-\tilde{\xi}_k/T}\right)\nonumber\\
+\frac{T}{2N}\sum_{K}\ln \left[1+g\Pi(K)\right],
\end{eqnarray}
where the `four-vector' $K=(\omega_k, {\bf k})$ stands for the bosonic Matsubara frequency and wave-vector. Here, $\tilde{\xi}_k=\varepsilon_k-\tilde{\mu}$, $\varepsilon_k=\hbar^2k^2/2m$ is the free-particle dispersion and note that in both sums the terms with ${\bf k}=0$ are omitted. Additionally, we have introduced the polarization operator $\Pi(K)=\Pi_0(K)+\Pi_T(K)$, where 
\begin{eqnarray}\label{Pi_0}
\Pi_0(K)=\frac{n_0}{\tilde{\xi}_k-i\omega_k}+{\rm c.c.},
\end{eqnarray}
denotes the condensate contribution to the dynamic structure factor of ideal Bose gas and
\begin{eqnarray}\label{Pi_T}
\Pi_T(K)=\frac{T}{V}\sum_{K'}\frac{1}{\tilde{\xi}_{k'}-i\omega_{k'}}\frac{1}{\tilde{\xi}_{|{\bf k}'+{\bf k}|}-i\omega_{k'+k}},
\end{eqnarray}
represents the impact of density fluctuations of the thermally stimulated single-particle excitations. It should be understood that the structure of the higher-order terms in (\ref{Omega}) is clear \cite{Vakarchuk_12} and there are no principal problems to include them. In practice, however, the numerical computations of the beyond-$1/N$-terms to the thermodynamic characteristics are substantially complicated. 

Minimization of $\Omega$ with respect to parameters $\varphi_0$ and $\psi_0$ leave us with $\varphi_0=-ing$ (and appropriately $\tilde{\mu}=\mu-ng$) and
\begin{eqnarray}\label{mu}
\mu=ng+\frac{T}{NV}\sum_{K}g_K\frac{\tilde{\xi}_k+i\omega_k}{E^2_K+\omega^2_k},
\end{eqnarray}
where we have introduced the effective two-body potential $g_K\equiv g/[1+g\Pi_T(K)]$ induced by the density fluctuations of non-condensed particles and the frequency-dependent quantity $E^2_K=\tilde{\xi}^2_k+2n_0g_K\tilde{\xi}_k$ which coincides with the Bogoliubov spectrum at absolute zero. The Matsubara frequency sum in the above formula should be calculated with factor $e^{i\omega_k\tau}$ ($\tau \to +0$) and even after that the integral over the wave-vector is divergent. This divergence is caused by the replacement of the original two-body short-range potential by the $\delta$-function. Therefore, we additionally have to rewrite everywhere in Eq.~(\ref{mu}) the `bare' coupling constant $g$ via the $s$-wave scattering length $a$ in the adopted approximation
\begin{eqnarray}\label{g}
g=\frac{4\pi\hbar^2a}{m}+\left(\frac{4\pi\hbar^2a}{m}\right)^2\frac{1}{NV}\sum_{{\bf k}}\frac{1}{2\varepsilon_k}\pm \ldots
\end{eqnarray}
The subsequent substitution in Eq.~(\ref{mu}) guaranties finiteness of the chemical potential \cite{Hryhorchak_19}. Finally, by using the well-know identity which relates the derivative of the thermodynamic potential with respect to the chemical potential to the particle number, we obtain the condensate density in the adopted approximation
\begin{align}\label{n_0}
&n_0=n-\left(1-\frac{1}{N}\right)\frac{1}{V}\sum_{{\bf k}}n(\tilde{\xi}_k/T)
-\frac{1}{NV}\sum_{{\bf k}}\frac{\partial n(\tilde{\xi}_k/T)}{\partial \tilde{\xi}_k}\nonumber\\
&\times \Sigma^{(1)}_R(\tilde{\xi}_k,k)-\frac{T}{NV}\sum_{K}\frac{i\omega_k+\tilde{\xi}_k+n_0g_K}{E^2_K+\omega^2_k}.
\end{align}
Here, $n(x)=1/(e^x-1)$ is the Bose distribution; the last sum should be carried out with factor $e^{+i\omega_k0}$ and $\Sigma^{(1)}_R(\omega,k)$ denotes the real part (after analytical continuation in the upper complex half-plane $i\omega_k\to \omega+i0$) of the leading-order self-energy $\Sigma(K)=\Sigma^{(1)}(K)/N+\ldots$ of normal Green's function \cite{Hryhorchak_19}
\begin{eqnarray}\label{Sigma}
\Sigma^{(1)}(K)=\frac{T}{V}\sum_{K'} \frac{g_{K'}(\tilde{\xi}^2_{k'}+\omega^2_{k'})}{E^2_{K'}+\omega^2_{k'}}\frac{1}{\tilde{\xi}_{|{\bf k}'+{\bf k}|}-i\omega_{k'+k}}.
\end{eqnarray}
It should be noted that Eq.~(\ref{n_0}) for the condensate density can be directly obtained by means of Green's function technique \cite{Hryhorchak_19}, which proves the consistency of the large-$N$ expansion approach. Now the interaction-renormalized temperature of BEC transition $T_c$ can be found from the condition $n_0=0$ and is determined by the following relation: r.h.s. of Eq.~(\ref{n_0})$=0$.

The standard procedure to obtain the density of superfluid component is well-described in literature and lies in the calculation of response of the thermodynamic potential to a slow motion of the whole system with velocity ${\bf v}$. In general, this shift of $\Omega$ can be accounted by the gauge transformation of the initial bosonic fields $\psi_{\sigma}(x)\to e^{im{\bf v}{\bf r}/\hbar}\psi_{\sigma}(x)$, $\psi^*_{\sigma}(x)\to e^{-im{\bf v}{\bf r}/\hbar}\psi^*_{\sigma}(x)$ in action (\ref{S}) with the subsequent repetition of the calculation scheme (\ref{S_eff}), (\ref{Omega}). From the practical point of view the problem reduces to the replacement of original chemical potential and Matsubara frequencies in the thermodynamic potential
\begin{eqnarray}\label{Omega_v}
\Omega_{\bf v}=\Omega(\mu\to \mu-mv^2/2, \omega_k\to\omega_k+i\hbar {\bf vk}).
\end{eqnarray}
The effectiveness of the above-described prescription for the superfluid density calculation and for the evaluation of the exact identities relating macroscopic observables to parameters of the low-lying excitations was previously proven for one \cite{Pastukhov_q2D,Pastukhov_InfraredStr} and two-component \cite{Pastukhov_twocomp,Konietin} Bose systems. In general, the superfluidity in the $N$-component Bose mixture, due to the Andreev-Bashkin effect, is characterized by the $N\times N$ symmetric matrix $n^{\sigma \sigma'}_s$ of superfluid densities. In our case, however, by supposing the same velocity for all $N$ species we smear out the information about the occurrence of this drag effect providing the diagonal structure of matrix $n^{\sigma \sigma'}_s=\delta_{\sigma \sigma'}n_s$ with $N$ times degenerated eigenvalue.

The further consideration should not cause any principal complications and by the brute force expansion of $\Omega_{\bf v}$ at small ${\bf v}$ we obtain
\begin{eqnarray}
\Omega_{{\bf v}\to 0}=\Omega+NVn_smv^2/2+o(v^2).
\end{eqnarray}
where $\Omega$ is given by Eq.~(\ref{Omega}) and the calculations of superfluid density $n_s$ per each sort of bosons yield
\begin{align}\label{n_s}
& n_s=n-\left(1-\frac{1}{N}\right)\frac{1}{V}\sum_{{\bf k}}n(\tilde{\xi}_k/T)
-\frac{1}{NV}\sum_{{\bf k}}\frac{\partial n(\tilde{\xi}_k/T)}{\partial \tilde{\xi}_k}\nonumber\\
&\times \Sigma^{(1)}_R(\tilde{\xi}_k,k)-\frac{2T}{3NV}\sum_{K}\varepsilon_k\frac{E^2_K-\omega^2_k}{(E^2_K+\omega^2_k)^2}.
\end{align}
The latter formula together with the condensate density (\ref{n_0}) represent the main result of this article that will be used below for the numerical computations.
But before we proceed to numerics few remarks should be made. The first important thing one should keep in mind is that the condensate and superfluid densities given by formulae (\ref{n_0}), (\ref{n_s}) both have additional irrelevant terms of higher orders in powers of parameter $1/N$. Indeed, these two equations contain the ideal-gas dispersion $\tilde{\xi}_k$ with the $1/N$-corrected effective chemical potential $\tilde{\mu}$ [see Eq.~(\ref{mu})]. Therefore, terms of order unity in Eqs.~(\ref{n_0}), (\ref{n_s})
\begin{eqnarray*}
\frac{1}{V}\sum_{{\bf k}}n(\tilde{\xi}_k/T)=\frac{1}{V}\sum_{{\bf k}}n(\varepsilon_k/T)-\frac{1}{V}\sum_{{\bf k}}\frac{\partial n(\varepsilon_k/T)}{\partial \varepsilon_k}\tilde{\mu},
\end{eqnarray*}
should be expanded in the Taylor series up to the first order in $\tilde{\mu}$. By direct comparison of Eq.~(\ref{mu}) and Eq.~(\ref{Sigma}) it is easy to verify that $\tilde{\mu}=\Sigma^{(1)}(0)/N$, which not only ensures the fulfilment of the Hugengoltz-Pines theorem but also provides the infrared convergence of integrals for $n_0$ and $n_s$. Particularly, it means that at temperatures not too close to $T_c$ the difference $\Sigma^{(1)}_R(\varepsilon_k, k)/N-\tilde{\mu}$ (which is under the wave-vector integral in the second terms of (\ref{n_0}) and (\ref{n_s}) (after substitution of $\tilde{\mu}$) is proportional to $k^2$, while $k$ goes to zero. In all other terms in Eqs.~(\ref{n_0}), (\ref{n_s}), which are already of order $1/N$ one is free to replace $\tilde{\xi}_k$ by $\varepsilon_k$.

Secondly, it is clearly seen from the above expressions that the superfluidity in the leading order of the large-$N$ expansion vanishes with vanishing of a condensate, i.e., the temperature of superfluid transition coincides with the critical temperature of the BEC. This commonly accepted idea about the nature of superfluid transition in three-dimensional Bose systems is in contradiction with some effective field theory studies \cite{Cooper_etal}. To see the coincidence of the BEC and superfluid transition temperatures in our approach, it is enough to realise that $E_K\to \tilde{\xi}_k\to \varepsilon_k$ (when $n_0\to 0$) and that last terms in (\ref{n_0}) and (\ref{n_s}) are equal to each other in the vicinity of critical point.

The third feature of the whole previous analysis is that the thermodynamic characteristics of $N$-component bosons with the fully symmetric interaction are expressed (in the large-$N$ limit) as a systematic series expansion in {\it integer} powers of $1/N$.

\section{Numerical results and discussion}
Taking into account the aforementioned discussion we are now in position to present the graphical results of the numerical integration and Matsubara frequency summation. To improve the convergence of these computations we have used the peculiarities of the $1/N$-expansion. In particular, it is worth mention that in the leading order, this theory applied to Bose systems incorporates two main ingredients, namely, the Bogoluibov theory at very low temperatures and inclusion of the finite-temperature density fluctuation of the non-condensed particles in the simplest approximation. Close to the Bose-Einstein transition temperature these fluctuations are highly developed providing the non-trivial critical behavior of the system. In what follows for the numerical purposes, one has to single out the finite-temperature Bogoliubov condensate depletion in (\ref{n_0}) and the Landau expression (with Bogoliubov's spectrum) for the superfluid density in (\ref{n_s}) and analytically perform the summations over frequencies. Now, the summands in remaining terms are well-behaviored at ultraviolet region and can be easily computed by the numerical methods. 

Additionally, it is important to consider the zero-frequency terms explicitly because they are actually responsible for the non-analytic behavior of all thermodynamic functions at the critical point and therefore have to be discussed in more detail. From equations determining $n_0$ and $n_s$ we find out that first three terms in r.h.s. of Eqs.~(\ref{n_0}), (\ref{n_s}) vanish linearly in the vicinity of $T_c$, while the last ones demonstrate logarithmic nonanalicity. It is easy to obtain the coefficient standing in front of this log-linear term and hinting the universal power-law behavior
\begin{eqnarray}\label{n_0_cr}
\frac{n_0}{n}\propto \delta t-\frac{4}{\pi^2N} \delta t\ln \delta t+\ldots\simeq(\delta t)^{2\beta}, 
\end{eqnarray}
\begin{eqnarray}\label{n_s_cr}
\frac{n_s}{n}\propto \delta t-\frac{16}{3\pi^2N} \delta t\ln \delta t+\ldots\simeq(\delta t)^{2\beta_s}, 
\end{eqnarray}
where $\delta t=\frac{T_c-T}{T_c}$ is the dimensionless deviation from the critical temperature. Moreover, these estimations can be easily extended to a more general case of arbitrary dimension $D>2$, where the Bose condensation phenomenon occurs. All we need is to evaluate the small-$k$ behavior of the `thermal' part of the polarization operator at zero Matsubara frequency $\Pi_T(K)|_{\omega_k=0}\propto k^{D-4}$. Then the substitution in the last term of equations for $n_0$ and $n_s$ (here prefactor $2/3$ should be replaced by $2/D$) leaves us with the result for critical exponents calculated in the accepted approximation
\begin{eqnarray}\label{beta}
\beta=\frac{1}{2}-\frac{\sin\left(\frac{D-2}{2}\pi\right)\Gamma(D-2)}{2\pi N\Gamma^2\left(\frac{D}{2}\right)}, 
\end{eqnarray}
\begin{eqnarray}\label{beta_s}
\beta_s=\frac{1}{2}-\frac{2\sin\left(\frac{D-2}{2}\pi\right)\Gamma(D-2)}{\pi D N\Gamma^2\left(\frac{D}{2}\right)},
\end{eqnarray}
where $\Gamma(x)$ is the gamma function. These findings particularly confirm the exact scaling relation for the superfluid density, $n_s/n\propto (\delta t)^{2\beta-\nu\eta}$ \cite{Lifshitz}, where $\nu$ and $\eta$ are correlation length and Fisher exponents, respectively. Indeed, taking into account values of $\nu$ and $\eta$ calculated to leading order in $1/N$ \cite{Abe}, we obtain our estimation for $\beta_s$.

Having found out the slope of the condensate and superfluid density curves close to $T_c$ we may proceed by presenting the results of full numerical computations. Putting $N=1$ we have calculated, for three values of the gas parameter $a^{3}n=10^{-6}$, $10^{-4}$ and $10^{-2}$, the temperature dependence of both $n_s$ and $n_0$ which are depicted in Figs.~1-3. We also plotted (symbols) results of quantum Monte Carlo simulations taken from \cite{Pilati_thesis}.Note that temperature in MC data points as well as in our curves is measured in units of $T_c$, where $T_c$ in our case is the $1/N$-shifted \cite{Hryhorchak} critical temperature of the system, while positions of MC points are rescaled to the BEC transition temperature obtained \cite{Pilati} in simulations. Of course, there is some discrepancy (maximum $\sim 5\%$) between these transition temperatures, which however, cannot substantially affect the character of the calculated temperature dependencies.
\begin{figure}[h!]
	\centerline{\includegraphics
		[width=0.5\textwidth,clip,angle=-0]{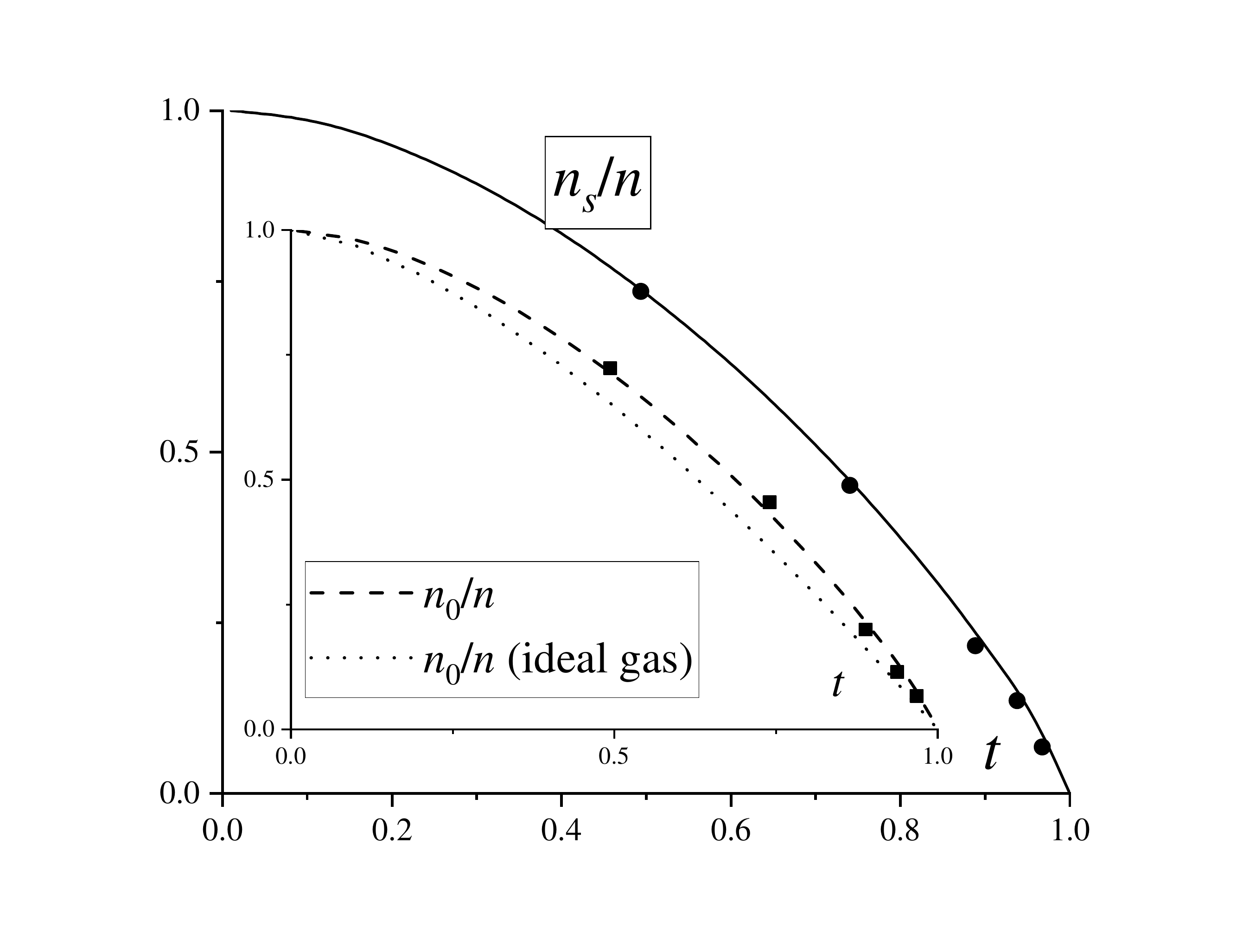}}
	\caption{The temperature dependence (in dimensionless units $t=T/T_c$) of the superfluid density for the Bose gas with $a^3n=10^{-6}$ (circles show the MC data). Inset: dashed and dotted lines correspond to the Bose condensate fraction of the interacting and ideal gases, respectively (squares denote MC points). Systematic errors of the MC simulations are typically smaller than symbol sizes.}
\end{figure}

The presented results of numerical calculations clearly demonstrate the natural tendency: in general, an increase of the interaction strength makes the coincidence with the MC data worse. Particularly, in Fig.~1 we observe a quite good matching in  whole temperature region. Regardless the fact that the interaction between particles is very weak the difference of calculated $n_0$ with the ideal Bose gas condensate density (dotted line), due to exceptional impact of the density fluctuation of the non-condensed particles, is visible. The curves in Fig.~1 should be compared to results \cite{Capogrosso-Sansone} obtained by means of the Beliaev technique extended on the finite-temperature region and combined with classical MC computations \cite{Prokofev}. In principle, for such weakly-interacting systems all approximate approaches \cite{Kita,Yukalovs_1,Watabe_13,Watabe_14,Yukalovs_2} work well and the only exception is the narrow neighborhood of the critical point, where results can vary. Even the standard Bogoliubov theory, which predicts the same critical temperature as in the ideal Bose gas, adequately describes the condensate depletion behavior of interacting particles at non-zero $T$ not too close to $T_c$ (at $a^3n=10^{-6}$ the Bogoliubov curve lies slightly $\sim 0.01$ above the calculated one).
\begin{figure}[h!]
	\centerline{\includegraphics
		[width=0.5\textwidth,clip,angle=-0]{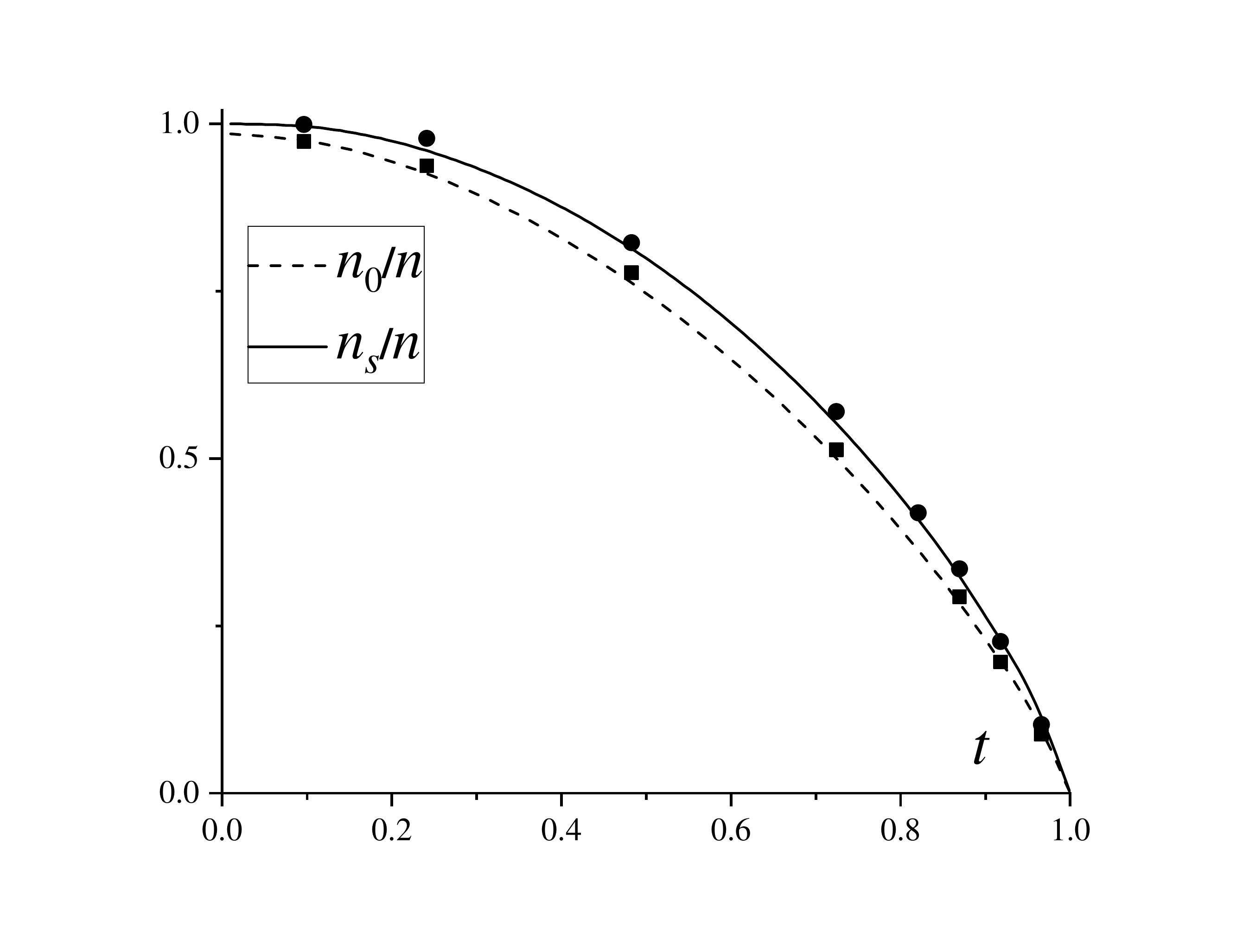}}
	\caption{The superfluid (solid line) and the Bose condensate (dashed line) densities for $a^3n=10^{-4}$ compared to MC results (circles for $n_s$ and squares for $n_0$).}
\end{figure}
\begin{figure}[h!]
	\centerline
	{\includegraphics
		[width=0.5\textwidth,clip,angle=-0]{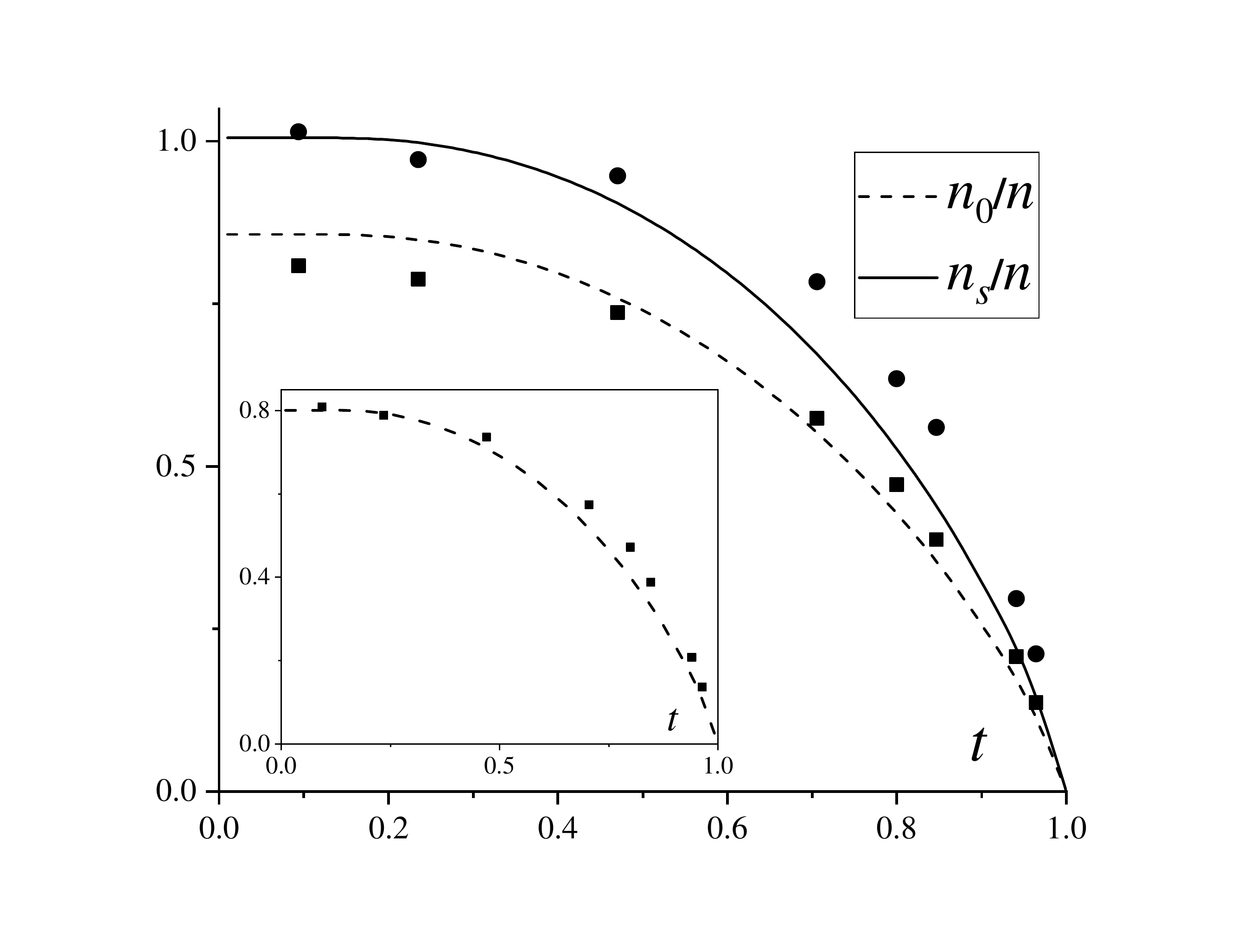}}
	\caption{Comparison of the temperature depletion of superfluid component (solid curve) and the Bose condensate (dashed curve) for $a^3n=10^{-2}$ to MC data (circles for $n_s$ and squares for $n_0$). The inset shows the dependence of $n_0$ (rescaled to the MC \cite{Giorgini} ground-state depletion) on temperature.}
\end{figure}

The quantitative agreement with MC results, as it is demonstrated in Figs.~2 and 3, is found to be worse with the growth of strength of the two-body repulsive potential. It is not surprising because at $a^3n=10^{-2}$ even the quantum depletion of the Bose condensate predicted by the large-$N$ expansion approach (which at $T=0$ is equivalent to the Bogoliubov theory) substantially differs from findings of the essentially exact MC simulations. The calculated temperature dependencies of the Bose condensate, however, are qualitatively correct. In order to show this we built in the inset of Fig.~3 the condensate density temperature behavior, rescaled to the zero-temperature value of $n_0$ obtained in MC \cite{Giorgini} simulations. A comparison of regimes of various coupling strengths for $n_s$ and $n_0$ leads us to conclusion that except for the ground-state depletion the temperature behavior of the Bose condensate is less sensitive to the interaction effects. The latter suggests that even for intermediate couplings the $1/N$ expansion in its simplest approximation is capable to capture the finite-temperature behavior of the Bose condensate density, while mechanisms responsible for the formation of the observed temperature dependence of the superfluid component in Bose systems with nonweak interparticle repulsion require further investigations.

\section{Conclusions}
In conclusion, we have shown that the combination of two simple ingredients, namely, the Bogoliubov theory at absolute zero and the inclusion of density fluctuations of the noncondensed particles at finite temperatures, that is incorporated in the simplest non-trivial approximation of the $1/N$-expansion method, is a promising tool for the quantitative study of the Bose-Einstein condensation phenomenon and emergence of superfluidity in Bose systems with short-range repulsion. Particularly, using the large-$N$ expansion to leading order we have calculated the temperature dependence of the superfluid and condensate densities for Bose system with point-like repulsive interaction between particles and demonstrated the efficiency of the presented approach by comparing our results to the Monte Carlo simulations data.

\begin{center}
{\bf Acknowledgements}
\end{center}
We are grateful to Prof.~Sebastiano~Pilati for providing us with the results of his Monte Carlo simulations. 
The author also thank Prof.~Andrij~Rovenchak for permanent help during the completion of this research.
This work was partly supported by Project FF-30F (No.~0116U001539) from the Ministry of Education and Science of Ukraine.


\begin{thebibliography}{99}
\bibitem{Kapitza} P.~Kapitza, Nature {\bf 141}, 74 (1938). 
\bibitem{Allen} J.~F.~Allen and A.~D.~Meissner, Nature {\bf 141}, 75 (1938).

\bibitem{Tisza} L.~Tisza, Nature {\bf 141}, 913 (1938).
\bibitem{Landau} L.~Landau, J.~Phys.~(U.S.S.R.) {\bf 5}, 71 (1947).

\bibitem{Pitaevskii} L.~P.~Pitaevskii, S.~Stringari, {\it Bose-Einstein condensation}
(Oxford University Press, Oxford, 2003).

\bibitem{Onofrio} R. Onofrio, C. Raman, J. M. Vogels, J. R. Abo-Shaeer, A. P. Chikkatur, and W. Ketterle, Phys.~Rev.~Lett. {\bf 85}, 2228 (2000).
\bibitem{Raman} C. Raman, R. Onofrio, J. M. Vogels, J. R. Abo-Shaeer, W. Ketterle,
J. Low Temp. Phys. {\bf 122}, 99 (2001).

\bibitem{Kashurnikov} V.~A.~Kashurnikov, N.~V.~Prokof'ev and B.~V.~Svistunov,
Phys.~Rev.~Lett. {\bf 87}, 120402 (2001).
\bibitem{Arnold} P.~Arnold and G.~Moore, Phys. Rev. Lett. {\bf 87}, 120401
(2001). 
\bibitem{Prokofev} N.~Prokof'ev, O.~Ruebenacker, and B.~Svistunov
Phys.~Rev.~A {\bf 69}, 053625 (2004).
\bibitem{Nho} K.~Nho and D.~P.~Landau, Phys.~Rev.~A {\bf 70}, 053614 (2004).
\bibitem{Floerchinger} S.~Floerchinger and C.~Wetterich, Phys.~Rev.~A {\bf 79}, 063602 (2009).
\bibitem{Isaule} F.~Isaule, M.~C.~Birse, and N.~R.~Walet, Phys.~Rev.~B {\bf 98}, 144502 (2018).

\bibitem{Cooper_etal} F.~Cooper, C.-C.~Chien, B.~Mihaila, J.~F. Dawson, and E.~Timmermans, Phys.~Rev.~Lett. {\bf 105}, 240402 (2010).
\bibitem{Kruglov} V.~I.~Kruglov, arXiv preprint arXiv:1511.00772.

\bibitem{Gaunt} A.~L.~Gaunt, T.~F. Schmidutz, I.~Gotlibovych, R.~P.~Smith, and Z.~Hadzibabic, Phys.~Rev.~Lett. {\bf 110}, 200406 (2013).
\bibitem{Lopes_etal} R.~Lopes, C.~Eigen, N.~Navon, D.~Cl\'ement, R.~P.~Smith, and Z.~Hadzibabic, Phys.~Rev.~Lett. {\bf 119}, 190404 (2017).
\bibitem{Ville} J.~L.~Ville, R.~Saint-Jalm, \'E.~Le~Cerf, M.~Aidelsburger, S.~Nascimb\`ene, J.~Dalibard, J.~Beugnon, Phys.~Rev.~Lett. {\bf 121}, 145301 (2018).


\bibitem{Andersen} J.~O.~Andersen, arXiv:cond-mat/0608265.
\bibitem{Chien} C.-C.~Chien, F.~Cooper, and E.~Timmermans, Phys.~Rev.~A {\bf 86}, 023634 (2012).
\bibitem{Diehl} H.~W.~Diehl and S.~B.~Rutkevich, Phys.~Rev.~E {\bf 95}, 062112 (2017).
\bibitem{Hryhorchak} O.~Hryhorchak and V.~Pastukhov, EPL (Europhysics Letters) {\bf 118}, 56003 (2017).
\bibitem{Hryhorchak_19} O.~Hryhorchak and V.~Pastukhov, J. Phys. A: Math. Theor. {\bf 52}, 025002 (2019).

\bibitem{Popov} V.~N.~Popov, {\it Functional Integrals and Collective Excitations} (Cambridge University Press, Cambridge, 1987).

\bibitem{Vakarchuk_12} I.~O.~Vakarchuk, V.~S.~Pastukhov, R.~O.~Prytula, Ukr.~J.~Phys. {\bf 57}, 1214 (2012).


\bibitem{Pastukhov_q2D} V.~Pastukhov, Ann. Phys. {\bf 372}, 149 (2016).
\bibitem{Pastukhov_InfraredStr} V.~Pastukhov, J.~Low Temp.~Phys. {\bf 186}, 148 (2017).
\bibitem{Pastukhov_twocomp} V.~Pastukhov, Phys.~Rev.~A {\bf 95}, 023614 (2017).
\bibitem{Konietin} P.~Konietin and V.~Pastukhov, J.~Low Temp.~Phys. {\bf 190}, 256 (2018).

\bibitem{Lifshitz} E.~M.~Lifshitz, L.~P.~Pitaevskii {\it Statistical Physics: Theory of the Condensed State (Pt. 2)} (Pergamon Press, Oxford, 1980)
\bibitem{Abe} R.~Abe and S.~Hikami, Prog.~Theor.~Phys.{\bf 49}, 442 (1973).

\bibitem{Pilati_thesis} S.~Pilati, PhD Thesis (Trento), unpublished.
\bibitem{Pilati} S.~Pilati, S.~Giorgini, and N.~Prokof'ev
Phys.~Rev.~Lett. {\bf 100}, 140405 (2008).

\bibitem{Capogrosso-Sansone} B.~Capogrosso-Sansone, S.~Giorgini, S.~Pilati, L.~Pollet, N.~Prokof'ev, B.~Svistunov, and M.~Troyer, New~J.~Phys. {\bf 12}, 043010 (2010).


\bibitem{Kita} T.~Kita, J.~Phys.~Soc.~Jap. {\bf 75}, 044603 (2006).
\bibitem{Yukalovs_1} V.~I.~Yukalov, E.~P.~Yukalova, Phys.~Rev.~A {\bf 76}, 013602 (2007).
\bibitem{Watabe_13} S.~Watabe and Y.~Ohashi, Phys.~Rev.~A {\bf 88}, 053633 (2013).
\bibitem{Watabe_14} S.~Watabe and Y.~Ohashi, Phys.~Rev.~A {\bf 90}, 013603 (2014).
\bibitem{Yukalovs_2} V.~I.~Yukalov, E.~P.~Yukalova, J. Phys. B: At. Mol. Opt. Phys. {\bf 47}, 095302 (2014).

\bibitem{Giorgini} S.~Giorgini, J.~Boronat, and J.~Casulleras
Phys.~Rev.~A {\bf 60}, 5129 (1999).




\end{thebibliography}
\end{document}